\documentclass[aps,prd,twocolumn,showpacs,floatfix]{revtex4}
\usepackage{exscale}
\usepackage{amsmath}
\usepackage[dvips]{graphicx}

\newcommand{\h}{\mathcal{H}}

\begin{document}

\title{Gravitational potential evolution in Unified Dark Matter Scalar Field Cosmologies: an analytical approach}

\author{Oliver F. Piattella$^{1,2}$}\email{oliver.piattella@gmail.com}\author{Daniele Bertacca$^{3,4,5}$}\email{bertacca@pd.infn.it}\email{daniele.bertacca@port.ac.uk}
\affiliation{$^{1}$Departamento de F\'isica, Universidade Federal do Esp\'irito Santo, avenida Ferrari 514, 29075-910 Vit\'oria, Esp\'irito Santo, Brazil}
\affiliation{$^2$INFN sezione di Milano, Via Celoria 16, I-20133 Milano, Italy}
\affiliation{$^{3}$Dipartimento di Fisica Galileo Galilei,  Universit\`{a} di Padova , via F. Marzolo, 8 I-35131 Padova, Italy}
\affiliation{$^4$INFN sezione di Padova, via F. Marzolo, 8 I-35131 Padova, Italy}
\affiliation{$^5$Institute of Cosmology \& Gravitation, University of Portsmouth,
Dennis Sciama Building, Portsmouth, PO1 3FX, United Kingdom}


\begin{abstract}
We investigate the time evolution of the gravitational potential $\Phi$ for a special class of non-adiabatic Unified Dark matter Models described by scalar field lagrangians. These models predict the same background evolution as in the $\Lambda$CDM and possess a non-vanishing speed of sound. We provide a very accurate approximation of $\Phi$, valid after the recombination epoch, in the form of a Bessel function of the first kind. This approximation may be useful for a future deeper analysis of Unified Dark Matter scalar field models.
\end{abstract}


\maketitle


\section{Introduction}

In the last decade the $\Lambda$CDM model has emerged as the {\it concordance} model of our universe. It assumes General Relativity (GR) as the underlying theory of gravity, and two unknown components dominating the late-times dynamics: Cold Dark Matter (CDM), responsible for structure formation, and a cosmological constant $\Lambda$ making up the balance for a spatially flat universe and driving the cosmic acceleration \cite{Riess:1998cb}. Issues (still open) with $\Lambda$, i.e. the {\it cosmological constant problem}, prompted the cosmologists' community to consider alternatives, mainly in the form of a dynamic component dubbed Dark Energy (DE) (see, for example, \cite{Tsujikawa:2010sc}). However, it should be recognized that, while some form of CDM is independently expected to exist within modifications or extensions of the Standard Model of high energy physics (for example, see \cite{Taoso:2007qk}), the really compelling reason to postulate DE has been the acceleration in the cosmic expansion \cite{Riess:1998cb}. It is mainly for this reason that it is worth investigating the hypothesis of CDM and DE being two aspects of a single Unified Dark Matter (UDM) component.

A large variety of UDM models, mainly based on adiabatic fluids or on scalar field Lagrangians have been investigated in the literature, see \cite{Bertacca:2010ct} for an up-to-date review. The most important feature appears to be the existence of pressure perturbations in the rest frame of the UDM fluid. The latter correspond to a Jeans length below which the growth of small density inhomogeneities is impeded and the evolution of the gravitational potential is characterized by an oscillatory and decaying behavior. Therefore, the general lesson that may be drawn is that a viable UDM model should be characterized by a small speed of sound (typically $\lesssim 10^{-4}c$ \cite{Camera:2010wm}), so that structure formation would not be hampered and the ISW effect signal would be compatible with CMB observation \cite{Bertacca:2007cv} (see also \cite{Sandvik:2002jz, Giannakis:2005kr, Piattella:2009kt, Bertacca:2010mt}).

Among those approaches which describe UDM models as scalar fields, we focus on the one presented in \cite{Bertacca:2008uf}. Here, the authors devise a reconstruction technique for models whose speed of sound is so small that the scalar field can cluster and, at the same time, the background expansion is identical to the one of the $\Lambda$CDM. This class of models has recently been investigated in \cite{Bertacca:2011in} by means of the cross-correlation between the ISW effect signal and the large scale structure (LSS) distribution and in \cite{Camera:2010wm} by means of parameters forecasts for future 3D cosmic shear surveys (see also \cite{Camera:2009uz}). In this brief report, we investigate in detail the evolution of the gravitational potential for this class of models. Through some simple mathematical manipulations we show how to cast the usual time evolution equation in a form which can be very reasonably approximated by a Bessel type equation, after the recombination epoch. The resulting solution provides a fitting function in excellent agreement with the numerical results.

Throughout the paper we use $8\pi G = c^2 = 1$ units and the $(+,-,-,-)$ signature for the metric. 


\section{The model and the basic equations}\label{Sec:Themodel}

Assuming a flat Friedmann-Lema\^{i}tre-Robertson-Walker (FLRW) background metric with scale factor $a(\eta)$ and $\eta$ the conformal time, the authors of \cite{Bertacca:2008uf} introduced a new class of non-adiabatic UDM scalar field models which, by allowing a pressure equal to $-\Lambda$ on cosmological scales, reproduce the same background expansion as the $\Lambda$CDM one. When the energy density of radiation becomes negligible, and disregarding also the baryonic component, the background evolution of the universe is completely described by
\begin{equation}\label{HLcdm}
\h^2 = H_0^2\left(\Omega_{\Lambda 0}a^2 + \Omega_{\rm m0}a^{-1}\right)\;,
\end{equation}
where $\h = a^{\prime}/a$ is the conformal time Hubble parameter and the prime denotes derivation with respect to the conformal time; $H_0$ is the Hubble constant and $\Omega_{\Lambda 0}$ and $\Omega_{\rm m0} =1 -\Omega_{\Lambda 0} $ are to be interpreted as the ``cosmological constant"   and ``dark matter'' density parameters, respectively. From the WMAP7 data: $\Omega_{\Lambda 0} = 0.734 \pm 0.029$ \cite{Komatsu:2010fb}. Moreover, the model is characterized by a speed of sound \cite{Garriga:1999vw, Mukhanov:2005sc} which has the following form \cite{Bertacca:2008uf}:
\begin{equation}\label{cs2eff}
 c_{\rm s}^2 = \frac{c_{\infty}^2}{1 + \left(1 - c_{\infty}^2\right)\nu \left(1 + z\right)^3}\;,
\end{equation}
where $c_{\infty}$ is its asymptotic ($a \to \infty$) limit and $\nu \equiv \Omega_{\rm m0}/\Omega_{\Lambda 0} \approx 0.362$.

Consider small inhomogeneities of the scalar field and write the perturbed FLRW metric in the longitudinal gauge \cite{Mukhanov:2005sc}: 
\begin{equation}
 ds^2 = a(\eta)^2\left[\left(1 + 2\Phi\right)d\eta^2 - \left(1 - 2\Phi\right)\delta_{ij}dx^idx^j\right]\;,
\end{equation}
where $\Phi = \Phi({\bf x}, \eta)$ is the gravitational potential. Considering plane-wave perturbations $\Phi \propto \exp{\left(i\bf{k}\cdot\bf{x}\right)}$ of comoving wave-number $k \equiv |\bf{k}|$ the evolution equation of $\Phi$ becomes
\begin{equation}\label{Phieq}
 \Phi'' + 3\h\Phi' + c_{\rm s}^2k^2\Phi + \left(2\h' + \h^2\right)\Phi = 0\;,
\end{equation}
where $c_{\rm s}^2$ is given in Eq.~\eqref{cs2eff}. Note that the gravitational potential evolves in the same way as in a $\Lambda$CDM universe for those modes with wavelengths larger than the sound horizon and/or when $c_{\rm s} \to 0$. 
In the same spirit of \cite{Giannakis:2005kr}, we express the gravitational potential $\Phi\left(k;\eta\right)$ of Eq.~\eqref{Phieq} as follows: 
\begin{equation}\label{HuGiannrel}
 \Phi\left(k;\eta\right) = \Phi_{\rm \Lambda}\left(k;\eta\right)T\left(k;\eta\right)\;,
\end{equation}
namely as a transfer function $T\left(k\;;\eta\right)$ applied to $\Phi_{\rm \Lambda}$, where $\Phi_{\rm \Lambda}$ is the $\Lambda$CDM gravitational potential, which solves Eq.~\eqref{Phieq} with $c_{\rm s} = 0$. In particular, we impose that $T(k;\eta)=1$ and  $dT(k;\eta)/d\eta=0$ for $\eta<\eta_{\rm rec}$, where $\eta_{\rm rec}$ is some epoch when the universe is matter dominated and the radiation component is negligible (usually later than the recombination epoch). Before this time, the sound speed must be very close to zero to allow for structure formation.

The knowledge of $T\left(k\;;\eta\right)$ is useful because it allows to directly translate informations, such as the total matter power spectrum, from the $\Lambda$CDM to the UDM models of \cite{Bertacca:2008uf}. In order to compare the predictions of the models with the observational data, one should express the density contrast as a function of the gravitational potential. For example, for scales smaller than the cosmological horizon and $z < z_{\rm rec}$ one has
\begin{equation}
\delta_{\rm {DM}}\left[k;\eta(z)\right] =\frac{-2k^{2}\Phi_{\rm \Lambda}\left[k;\eta(z)\right]T\left[k;\eta(z)\right]}{3H_0^2\Omega_{\rm m0}\left(1 + z\right)}\;,
\end{equation}
where $\delta_{\rm {DM}}$ is the UDM density contrast \cite{Piattella:2009kt, Bertacca:2011in}.


\section{The equation for $T$}\label{Sec:Transfun}

Inserting Eq.~\eqref{HuGiannrel} into Eq.~\eqref{Phieq} and using the equation for $\Phi_\Lambda$ one obtains
\begin{equation}\label{Teq}
 T'' + \left[2\frac{\Phi_\Lambda'}{\Phi_\Lambda} + 3\mathcal{H}\right]T' + c_{\rm s}^2k^2T = 0\;.
\end{equation}
Being such equation not analytically solvable, in \cite{Bertacca:2011in} the authors employed the following approximation for $T$:
\begin{equation}\label{fittingfunc}
 T \approx j_{0}\left[\mathcal{A}(\eta)k\right]\;, \qquad \mathcal{A}(\eta) \equiv \int^{\eta}_{\eta_{\rm rec}}c_{\rm s}(\eta')d\eta'\;.
\end{equation}
The integral defining $\mathcal{A}(\eta)$ is not solvable, but since we are going to consider $c_{\infty}^2 \leq 10^{-2}$ \cite{Bertacca:2007cv, Bertacca:2011in}, it is convenient to expand $c_{\rm s}$ in Taylor series near $c_{\infty} = 0$:
\begin{equation}\label{cs2eff1storder}
c_{\rm s} = \frac{c_{\infty}}{\sqrt{1 + \nu \left(1 + z\right)^3}} + O\left(c_{\infty}^3\right)\;.
\end{equation}
Retaining just the first order of this expansion, plugging it into the formula for $\mathcal{A}(\eta)$ in Eq.~\eqref{fittingfunc} and changing the integration variable to the scale factor $a$, one finds
\begin{equation}\label{Aformula2}
 \mathcal{A} \approx c_{\infty}\int^{a}_{a_{\rm rec}}\frac{da}{\h a\sqrt{1 + \nu a^{-3}}} = \frac{c_{\infty}}{H_0\Omega_{\Lambda}^{1/2}}\int^{a}_{a_{\rm rec}}da\frac{a}{a^3 + \nu}\;.
\end{equation}
For simplicity, choosing as lower integration limit $a = 0$, the integration can be performed analytically and the result is:
\begin{eqnarray}\label{Aformula3}
 \frac{H_0\Omega_{\Lambda}^{1/2}\mathcal{A}}{c_{\infty}} &\approx& \frac{\sqrt{3}\pi}{18\nu^{1/3}} + \frac{1}{6\nu^{1/3}}\ln\frac{a^2 - a\nu^{1/3} + \nu^{2/3}}{\left(a + \nu^{1/3}\right)^2} \nonumber\\ &+& \frac{\sqrt{3}}{3\nu^{1/3}}\arctan\left(\frac{\sqrt{3}}{3}\frac{2a - \nu^{1/3}}{\nu^{1/3}}\right)\;.
\end{eqnarray}
As graphically showed in \cite{Bertacca:2011in}, Eq.~\eqref{fittingfunc} together with Eq.~\eqref{Aformula3} gives quite an accurate approximation of $T$ for this class of UDM models. In the following, we give a motivation for such good agreement and our discussion shall allow to find an even more accurate fitting function. In the next section, we analyze the term $\Phi_\Lambda'/\Phi_\Lambda$, which plays an important role in Eq.~\eqref{Teq}.


\section{The $\Lambda$CDM gravitational potential}\label{Sec:LCDMgravpot}

In terms of the scale factor $a$, the evolution equation for $\Phi_\Lambda$ becomes
\begin{equation}\label{PhiLa}
 \frac{d^2{\Phi}_{\Lambda}}{da^2} + \left[\frac{4}{a} + \frac{\mathcal{H}'}{a\mathcal{H}^2}\right]\frac{d\Phi_{\Lambda}}{da} + \left[\frac{1}{a^2} + \frac{2\mathcal{H}'}{a^2\mathcal{H}^2}\right]\Phi_{\Lambda} = 0\;.
\end{equation}
Considering $a \rightarrow \xi =(1-\sqrt{a^3/\nu + 1})/2$, we cast Eq.~\eqref{PhiLa} in the form of a hypergeometric equation:
\begin{equation}\label{PhiLahypergeom}
 \xi\left(\xi - 1\right)\dfrac{d^2\Phi_{\Lambda}}{d\xi^2} + \frac{11}{6}\left(2\xi - 1\right)\frac{d\Phi_{\Lambda}}{d\xi} + \frac{4}{3}\Phi_{\Lambda} = 0\;.
\end{equation}
Choosing $\Phi_{\Lambda}(k;a_{\rm rec}) = 1$ and $\left. d\Phi_\Lambda/da\right|_{a_{\rm rec}} = 0$ as normalized initial conditions, and applying a suitable Kummer transformation \cite{AS} one can cast the solution in the following form:
\begin{equation}\label{PhiLsol}
 \Phi_{\Lambda}\left(k;a\right) = \Phi\left(k;0\right)T_{\rm m}\left(k\right) {}_{2}F_{1}\left(\frac{1}{3},1;\frac{11}{6};-\frac{a^3}{\nu}\right)\;,
\end{equation}
where $\Phi\left(k;0\right)$ is the primordial gravitational potential at large scales, set during inflation \cite{Dodelson:2003ft}, and $T_{\rm m}\left(k\right)$ is the matter transfer function up to recombination suggested, for example, by \cite{Bardeen:1985tr} or \cite{Eisenstein:1997ik}
(for a more general and deeper analysis of the $\Lambda$CDM model, see also \cite{Hu:1998tj}). From \eqref{PhiLsol}, we are able to compute the term $\Phi_\Lambda'/\Phi_\Lambda$ in Eq.~\eqref{Teq}.


\section{The fitting function}\label{Sec:fittfun}

\begin{figure}[htbp]
\begin{center}
\includegraphics[width=\columnwidth]{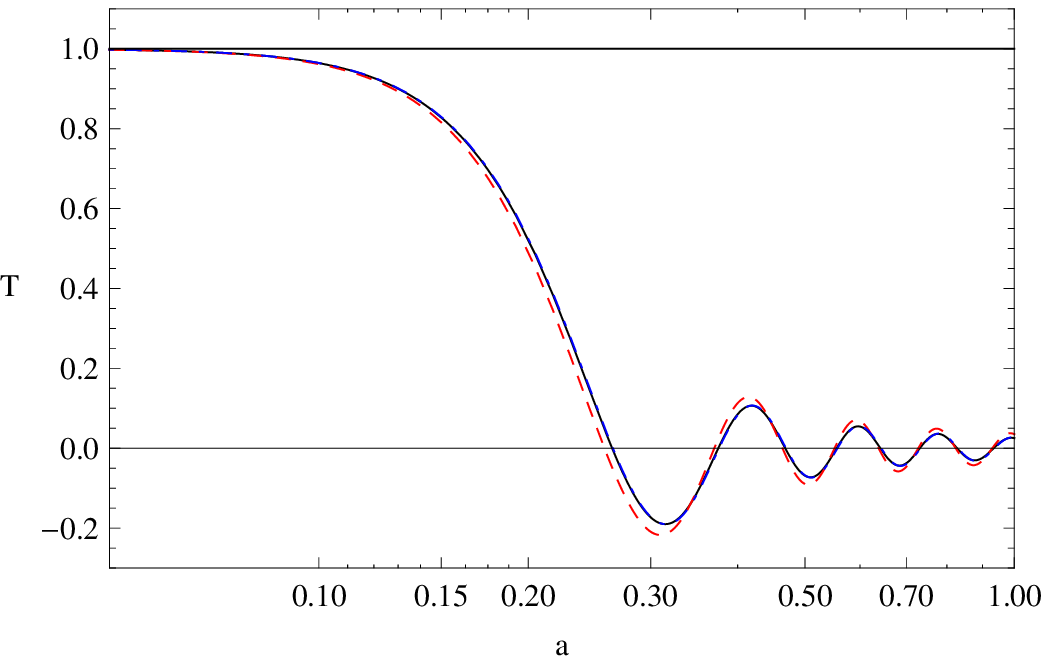}\\
\includegraphics[width=\columnwidth]{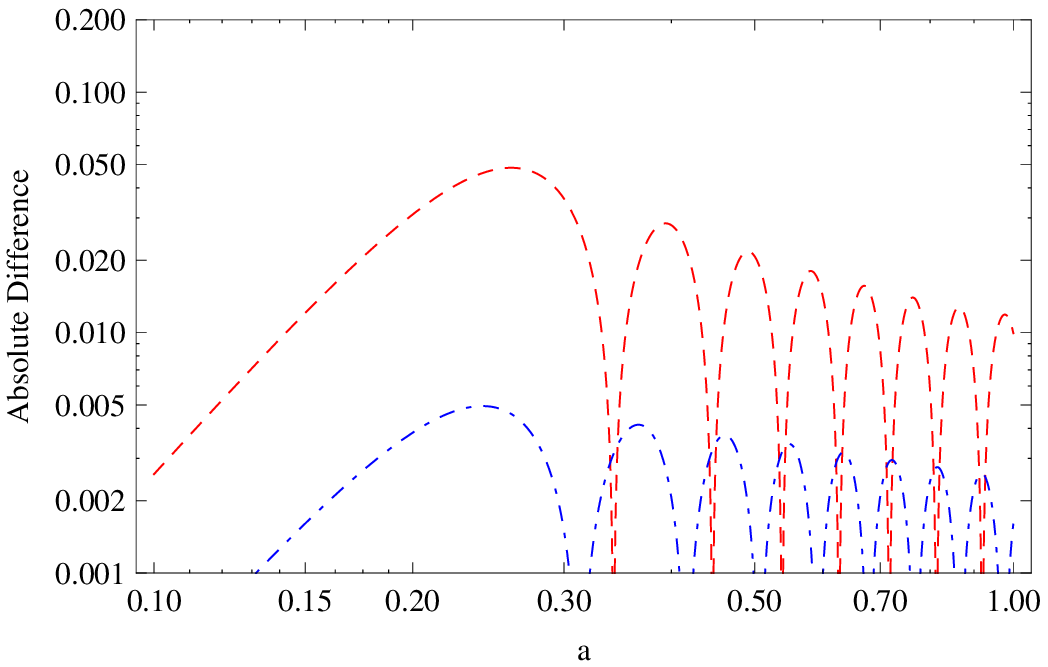}\\
\caption{(Color online). Upper panel: comparisons for $c_{\infty}^2 = 10^{-2}$ and $k = 0.1$ $h$ Mpc$^{-1}$ of the numerical result (black solid line) for $T$, the fitting function \eqref{fittingfunc} (red dashed line) and the fitting function \eqref{Teqg94sol2} (blue dash-dotted line). Lower panel: absolute differences between the numerical result for $T$ and the fitting function \eqref{fittingfunc} (red dashed line) and the fitting function \eqref{Teqg94sol2} (blue dash-dotted line).}
\label{comp1}
\end{center}
\end{figure}

\begin{figure}[htbp]
\begin{center}
\includegraphics[width=\columnwidth]{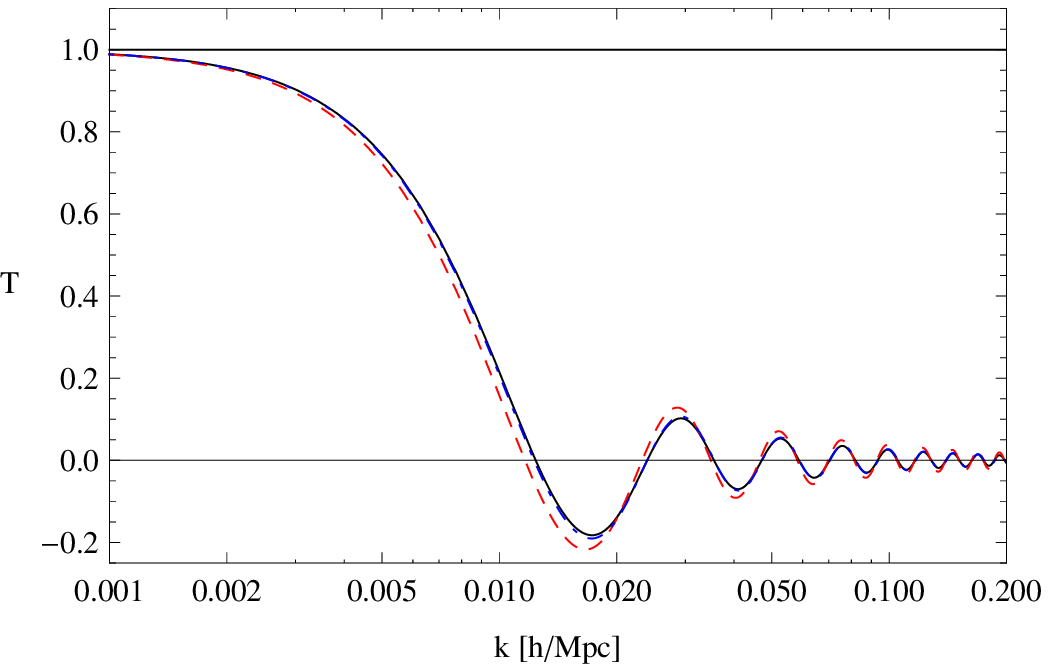}\\
\includegraphics[width=\columnwidth]{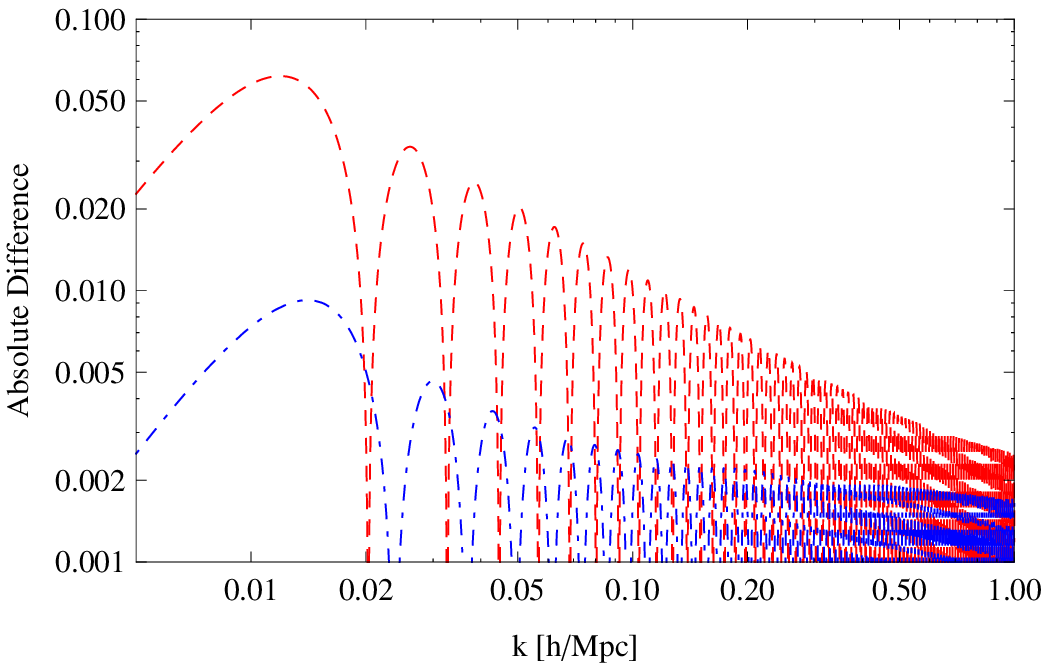}\\
\caption{(Color online). Upper panel: comparisons for $c_{\infty}^2 = 10^{-2}$ and $a = 1$ of the numerical result (black solid line) for $T$, the fitting function \eqref{fittingfunc} (red dashed line) and the fitting function \eqref{Teqg94sol2} (blue dash-dotted line). Lower panel: absolute differences between the numerical result for $T$ and the fitting function \eqref{fittingfunc} (red dashed line) and the fitting function \eqref{Teqg94sol2} (blue dash-dotted line).}
\label{comp2}
\end{center}
\end{figure}

Introducing a convenient new variable  $g(\eta,k) \equiv k\mathcal{A}(\eta)$, Eq.~(\ref{Teq}) becomes
\begin{equation}\label{Teqg}
g^2\; \frac{d^2T}{dg^2}+g\;b(g,\eta)\frac{dT}{dg}+ g^2\;T = 0\;,
\end{equation}
where
\begin{equation}\label{bcoeff}
b(g,\eta)\equiv \frac{g''g}{g^{'2}} + \frac{g}{g'}\left(2\frac{\Phi_\Lambda'}{\Phi_\Lambda} + 3\mathcal{H}_{\Lambda}\right)\;.
\end{equation}
\begin{figure}[htbp]
\begin{center}
\includegraphics[width=\columnwidth]{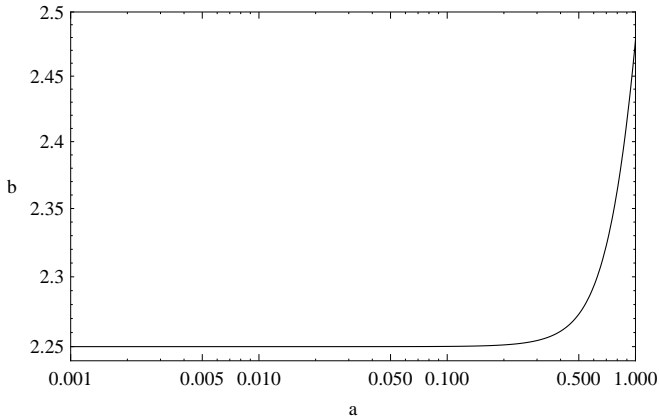}\\
\caption{Plot of $b$ as function of the scale factor $a$ and for $\nu = 0.362$.}
\label{figbxc}
\end{center}
\end{figure}
\newline
In Fig. \ref{figbxc} we plot $b$ as function of the scale factor and for $\nu = 0.362$. Notice that $b \approx 9/4$ for most of the evolution, i.e. up to $a \approx 0.5$. Then, it varies from $9/4$ ($2.25$) to $2.47$ at $a = 1$. By approximating $b = 9/4$ we make a relative error of about 9\% in $a = 1$. Remarkably, the choice $b = 9/4$ renders Eq.~(\ref{Teqg}) a Bessel-type equation \cite{AS}, whose solution is
\begin{equation}\label{Teqg94sol2}
 T(g) = 2^{5/8}\Gamma\left(\frac{13}{8}\right)g^{-\frac{5}{8}}J_{\frac{5}{8}}(g)\;.
\end{equation}
In Figs.~\ref{comp1} and \ref{comp2}, the fitting function \eqref{Teqg94sol2} is compared with the fitting function of Eq.~\eqref{fittingfunc}, the same adopted in \cite{Bertacca:2011in}. For both the cases, the functional form of $\mathcal{A}(\eta)$ is the approximated one given in Eq.~\eqref{Aformula3}. We choose, as a representative example, the case $c_{\infty}^2 = 10^{-2}$, for which the oscillatory behavior is most pronounced.

The reason for the success of $b = 9/4$ is the following: on average, in the interval $a\in(10^{-3},1)$ we do have $b \approx 9/4$. Indeed, computing the average of $b$ with respect to $a$ or $g$, we approximatively obtain $\langle b \rangle_a \approx  \langle b \rangle_g \approx 2.30$, where 
\begin{equation}
\langle b \rangle_x \equiv \int_{x_{\rm rec}}^{x_0}\frac{b(\tilde{x})d\tilde{x}}{x_0-x_{\rm rec}}\;.
\end{equation}

\begin{figure}[htbp]
\begin{center}
\includegraphics[width=\columnwidth]{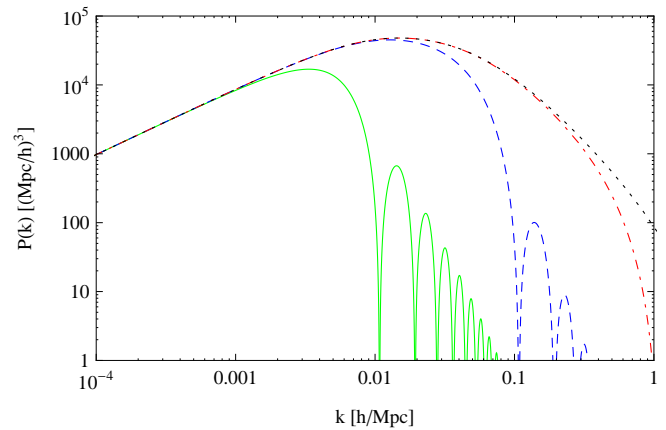}\\
\caption{(Color online). Power spectrum $P(k)$ of the UDM clustering energy density component $\delta_{\rm DM}$. From left to right, $c_{\infty}^2 = 10^{-2}$, $10^{-4}$, $10^{-6}$, $10^{-7}$ (green solid, blue dashed, red dot-dashed, black dotted lines, respectively).}
\label{Pk}
\end{center}
\end{figure}

In conclusion, from Eq.\ (\ref{Teqg94sol2}) and the definition of $\delta_{\rm {DM}}\left(k;z\right)$, in Fig.\ \ref{Pk} we show the power spectrum $P(k)$ of the UDM energy density component that clusters:
\begin{equation}
\langle \delta_{\rm {DM}}\left({\bf k};z=0\right)  \delta_{\rm {DM}}\left({\bf k'};z=0\right)  \rangle = (2\pi)^3 \delta({\bf k}+{\bf k}') P(k)\;,
\end{equation}
see also \cite{Bertacca:2011in}.


\section{Conclusions}\label{Sec:conclusions}

We have investigated in detail the evolution of the gravitational potential for a class of non-adiabatic Unified Dark Matter models which provide a background evolution similar to the $\Lambda$CDM one and possess a non-vanishing speed of sound. Such class of models has been investigated in \cite{Bertacca:2011in} where a convenient fitting function for the gravitational potential, in the form of a spherical Bessel function of order zero, has been employed. In particular, we have analyzed the evolution equation of a suitable transfer function $T\left(k\;;\eta\right)$, defined in \eqref{HuGiannrel}, introducing the new variable $g = k\mathcal{A}(\eta)$, cf. Eq.~\eqref{Teqg}, and obtaining a new fitting function, see Eq.~\eqref{Teqg94sol2}, that describes with great accuracy the evolution of $T\left(k\;;\eta\right)$ after the recombination epoch. This fitting function provides a useful tool for testing this class of models, especially in the light of new data on the cosmic background radiation and the weak gravitational lensing coming from, among the others, the Planck collaboration \cite{Planck} and the Euclid project \cite{Refregier:2010ss}.

A future development of our work would be to investigate in more detail the evolution of the gravitational potential in UDM models characterized by a fast transition \cite{Piattella:2009kt, Bertacca:2010mt}. In the adiabatic case, the speed of sound is a function highly peaked in the moment of the transition, being exponentially vanishing elsewhere. We expect the analytic approximate treatment to be more complicated than the one here presented, because it would involve mathematical techniques for dealing with rapidly varying coefficient functions in a second order differential equation.


\section{Acknowledgements}

OFP was supported by the CNPq (Brazil) contract 150143/2010-9.
DB would like to acknowledge the ICG (Portsmouth) for the hospitality during the development of this project and La ``Fondazione Ing. Aldo Gini" for support. DB research has been  partly supported by ASI  contract I/016/07/0 ``COFIS". The authors also thank  N.\ Bartolo, S.L Cacciatori, J.C. Fabris, S.\ Matarrese, and A.\ Raccanelli for discussions and suggestions.


\bibliographystyle{plain}

\end{document}